\documentclass[aps,prb,twocolumn,letterpaper,superscriptaddress]{revtex4}

\usepackage{amsmath,amssymb,amsfonts,amsthm}
\usepackage{graphicx,bm}

\renewcommand{\vec}[1]{\bm{#1}}

\begin{document}

\title{Towards a microscopic theory of toroidal moments in bulk periodic
crystals}

\date{\today}

\author{Claude Ederer}
\affiliation{Department of Physics, Columbia University, 538 West
120th Street, New York, NY 10027, U.S.A.}
\email{ederer@phys.columbia.edu}
\altaffiliation[Present address: ]{School of Physics, Trinity College,
  Dublin 2, Ireland}
\author{Nicola A.~Spaldin}
\affiliation{Materials Department, University of California, Santa
  Barbara, CA 93106-5050, U.S.A.}
 
\begin{abstract}
We present a theoretical analysis of magnetic toroidal moments in
periodic systems, in the limit in which the toroidal moments are
caused by a time and space reversal symmetry breaking arrangement of
localized magnetic dipole moments. We summarize the basic definitions
for finite systems and address the question of how to generalize these
definitions to the bulk periodic case. We define the
\emph{toroidization} as the toroidal moment per unit cell volume, and
we show that periodic boundary conditions lead to a multivaluedness of
the toroidization, which suggests that only differences in
toroidization are meaningful observable quantities. Our analysis bears
strong analogy to the ``modern theory of electric polarization'' in bulk
periodic systems, but we also point out some important differences
between the two cases. We then discuss the instructive example of a
one-dimensional chain of magnetic moments, and we show how to properly
calculate changes of the toroidization for this system. Finally, we
evaluate and discuss the toroidization (in the local dipole limit) of
four important example materials: BaNiF$_4$, LiCoPO$_4$, GaFeO$_3$,
and BiFeO$_3$.
\end{abstract}


\maketitle

\section{Introduction}
\label{intro}

\begin{figure}
\centerline{\includegraphics[width=0.95\columnwidth]{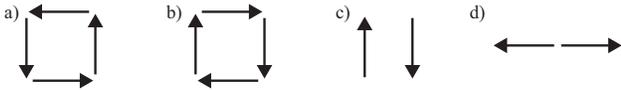}}
\caption{Simple arrangements of magnetic moments which can lead to
  toroidal moments. (a) and (b) have equal and opposite toroidal
  moments. The antiferromagnetic arrangement in (c) has a toroidal
  moment, whereas that in (d) does not.}
\label{toroidal_cartoons}
\end{figure}

The recent resurgence of interest in magneto-electric multiferroics
has prompted discussion of the relevance of the concept of magnetic
toroidal moments in such systems (see
e.g. Refs.~\onlinecite{Schmid:2001,Schmid:2003,Schmid:2004,Fiebig:2005,Arima_et_al:2005,Sawada/Nagaosa:2005,VanAken_et_al:2007}).
A magnetic toroidal moment is represented by a time-odd polar (or
``axio-polar'', see Ref.~\onlinecite{Ascher:1974}) vector, which
changes sign under both time inversion and space inversion, and is
generally associated with a ``circular'' or ``ring-like'' arrangement
of spins (see Fig.~\ref{toroidal_cartoons} for some
examples).\cite{Dubovik/Tugushev:1990} Materials in which the toroidal
moments are aligned cooperatively -- so-called {\it ferrotoroidics} --
have been proposed to complete the group of primary
ferroics.\cite{Schmid:2003,Schmid:2004,VanAken_et_al:2007} Further
interest stems from the fact that the toroidal moment is related to
the antisymmetric part of the linear magneto-electric
tensor;\cite{Gorbatsevich/Kopaev/Tugushev:1983,Sannikov:1997} this
points to an important role played by the magnetic toroidal moment for
magneto-electric coupling phenomena.

Magnetic toroidal moments in condensed matter systems were first
studied in the former Soviet Union during the 1980s, mostly in the
context of the so-called ``excitonic insulator'' model (see
Refs.~\onlinecite{Dubovik/Tugushev:1990,Gorbatsevich/Kopaev:1994} for
reviews of this work). At about the same time, Sannikov and Zheludov
proposed the toroidal moment as the primary order parameter for the
low temperature phase transition in multiferroic nickel iodine
boracite.\cite{Sannikov/Zheludov:1985} Since in these early studies
the toroidal moment was mostly treated as a macroscopic order
parameter, no particular attention was paid to the peculiarities
arising from the microscopic definition of the toroidal moment. It is
therefore the purpose of the present paper to give a detailed analysis
of the properties of toroidal moments starting from the
\emph{microscopic} definition and focusing especially on effects
resulting from the periodic boundary condition in crystalline
solids. In particular, we address the following questions:
\begin{enumerate}
\item{How should the toroidal moment density, or {\it toroidization},
of a bulk periodic solid be formally defined?}
\item{Is there a consistent way to treat the origin dependence of the 
toroidal moment?}
\item{What are the consequences of the periodic boundary conditions
within a bulk crystalline solid?}
\end{enumerate}
In addition, we apply our newly developed concepts to evaluate and
analyze the toroidization of four example materials: the
antiferromagnetic ferroelectrics BaNiF$_4$ and BiFeO$_3$, the polar
ferrimagnet GaFeO$_3$, and the strongly magneto-electric material
LiCoPO$_4$.

To avoid confusion, we point out that there also has been some recent
discussion about {\it electric} toroidal moments $\vec{g}$, defined as
$\vec{g} = \frac{1}{2} \sum_i \vec{r}_i \times \vec{p}_i$ where
$\vec{p}_i$ is the local electric dipole moment at position
$\vec{r}_i$ and the summation extends over all dipole moments in the
system.\cite{Dubovik/Tugushev:1990,Naumov/Bellaiche/Fu:2004,Prosandeev_et_al:2006}
The vector $\vec{g}$ is fundamentally different from the magnetic
toroidal moment, since it is both time- and space-inversion
symmetric. It has been used to characterize circular domains in
nano-scale ferroelectric
materials.\cite{Naumov/Bellaiche/Fu:2004,Prosandeev_et_al:2006} In the
following we exclusively discuss the case of \emph{magnetic} toroidal
moments, i.e. the term ``toroidal moment'' is always used in the sense
of ``magnetic toroidal moment''. We point out, however, that some of
our general considerations regarding origin-dependence and the effect
of periodic boundary conditions are applicable to the case of electric
toroidal moments as well.

This paper is organized as follows. We begin by summarizing some of
the basic definitions and then discuss the limit where the toroidal
moment is caused by a time and space reversal symmetry breaking
arrangement of localized magnetic moments (Sec.~\ref{definitions}). In
Sec.~\ref{origin} we then analyze the origin dependence of the
toroidal moment by decomposing the magnetic moment distribution into a
fully compensated, generally non-collinear, antiferromagnetic part and
a non-compensated, collinear, ferromagnetic part. In
Sec.~\ref{sec:bulk} we define the toroidization as the toroidal moment
per unit cell volume, and we show that the periodic boundary
conditions lead to a multivaluedness of the toroidization, similar to
the case of the electric polarization in bulk periodic solids (see
Refs.~\onlinecite{King-Smith/Vanderbilt:1993,Vanderbilt/King-Smith:1993,Resta:1994}).
This multivaluedness suggests that, as in the case of the electric
polarization, only differences in toroidization can be physically
observable quantities. In Sec.~\ref{chain} we illustrate some
consequences of the periodicity by using the example of a
one-dimensional antiferromagnetic chain of magnetic moments. In
Sec.~\ref{examples} we evaluate the toroidizations of four example
materials: BaNiF$_4$, LiCoPO$_4$, GaFeO$_3$, and BiFeO$_3$. Finally,
in Sec.~\ref{conclusions} we summarize our main conclusions and
discuss the correspondence between the microscopic toroidal moment
described in this paper and some phenomenological quantities with the
same symmetry, that have recently appeared in the multiferroics
literature.

\section{Definitions}
\label{definitions}

The toroidal moment $\vec{t}$ corresponding to a current density
distribution $\vec{j}(\vec{r})$ is defined as (see
Ref.~\onlinecite{Dubovik/Tugushev:1990}):
\begin{equation}
\vec{t} = \frac{1}{10c} \int \left( \vec{r} (\vec{r}\cdot\vec{j}) - 2 r^2
  \vec{j} \right) d^3r \quad ,
\label{t_formal}
\end{equation}
where $c$ indicates the speed of light in vacuum. The toroidal moment
in the form of Eq.~(\ref{t_formal}) emerges from the multipole
expansion of an arbitrary localized current
distribution.\cite{Dubovik/Tugushev:1990} Its physical significance
can be seen by noting that Eq.~(\ref{t_formal}) is identically
satisfied by the current distribution
\begin{equation}
\vec{j}(\vec{r}) = c \nabla \times \nabla \times \delta(\vec{r})
\vec{t} \quad ,
\label{tj}
\end{equation}
which represents an elemental toroidal moment centered at the
origin. Evaluating the interaction energy $E$ of this current density
with the electromagnetic field $\vec{A}(\vec{r})$ yields (after
partial integration):
\begin{equation}
E = - \frac{1}{c} \int \vec{j}(\vec{r}) \cdot \vec{A}(\vec{r}) \,
  d^3r = - \vec{t} \cdot \nabla \times \vec{B}(0) \quad ,
\label{Eoft}
\end{equation}
where $\vec{B}(0) = \nabla \times \vec{A}(\vec{r})\vert_{\vec{r}=0}$
is the magnetic field at the site of the toroidal moment. From
Eq.~(\ref{Eoft}) it can be seen that the toroidal moment couples to
the curl of the magnetic field such that a toroidal system in a
magnetic field has lowest energy when its toroidal moment is aligned
parallel to the curl of the magnetic field.

The definition of the toroidal moment can be re-cast into a more
convenient form, by noting that the current vector can be decomposed
into longitudinal ($\nabla \times \vec{j}_\parallel=0$) and
transversal ($\nabla \cdot \vec{j}_\perp=0$) parts. The longitudinal
part of $\vec{j}(\vec{r})$ is related to time derivatives of the
charge multipole moments through the continuity equation $\dot{\rho} +
\nabla \cdot \vec{j} = 0$, and does not contribute to the toroidal
moment.\cite{Dubovik/Tugushev:1990} The transverse part of the current
density $\vec{j}_\perp(\vec{r})$ can be written as the curl of the
\emph{magnetization density}
$\vec{\mu}(\vec{r})$:\cite{footnote:decomp}
\begin{equation}
\vec{j}_\perp(\vec{r}) = c \nabla \times \vec{\mu}(\vec{r}) \quad .
\label{j_perp}
\end{equation}
Inserting Eq.~(\ref{j_perp}) in Eq.~(\ref{t_formal}) gives the
toroidal moment in terms of
$\vec{\mu}(\vec{r})$:\cite{footnote:mu_trans}
\begin{equation}
\label{t_mu}
\vec{t} = \frac{1}{2}\int \vec{r} \times \vec{\mu}(\vec{r}) \, d^3r
\quad .
\end{equation}
While in principle, for a finite system with known magnetization
density, this expression can be used to evaluate the toroidal moment,
it is not directly applicable to extended systems where periodic boundary
conditions are employed. The difficulties resemble those encountered
in early attempts to calculate the electric polarization (see
Ref.~\onlinecite{Resta:1994}): for a general continuous magnetization
density $\vec{\mu}(\vec{r})$, Eq.~(\ref{t_mu}) evaluated over one unit
cell will lead to arbitrary values, depending on the special choice
of unit cell used in the calculation.

In the case of the electric polarization, a general solution to this
problem is achieved by evaluating the electric polarization directly
from the electronic wave-function using Wannier
representations.\cite{King-Smith/Vanderbilt:1993,Vanderbilt/King-Smith:1993,Resta:1994}
In principle, a similar approach seems appropriate for the case of the
toroidal moment. In the present paper, we pursue a somewhat simpler
yet rather instructive approach by assuming that the magnetization
density can be well represented by a distribution of localized
magnetic moments $\{\vec{m}_\alpha\}$ at sites $\vec{r}_\alpha$:
\begin{equation}
\label{spin_dist}
\vec{\mu}_\text{loc}(\vec{r}) = \sum_\alpha \vec{m}_\alpha
\delta(\vec{r} - \vec{r}_\alpha) \quad .
\end{equation}
This results in the toroidal moment:
\begin{equation}
\label{t_spin}
\vec{t} = \frac{1}{2} \sum_\alpha \vec{r}_\alpha \times \vec{m}_\alpha
\quad .
\end{equation}
The simplification to localized magnetic moments avoids the technical
difficulties of a full gauge invariant wave-function formulation but
retains all the peculiarities resulting from the periodic boundary
conditions within bulk systems. It also allows a more intuitive
analysis of several important prototype systems. Our study thus
represents a first step towards a full microscopic theory of toroidal
moments in crystalline solids and can be used as the basis for future
developments.

We note however, that in some cases the restriction to localized
magnetic moments can represent a severe simplification. Obviously, the
local moment picture neglects the possibility of toroidal moments
arising from non-localized magnetization densities, but in addition
the symmetry of the magnetic moment distribution in
Eq.~(\ref{spin_dist}) can eventually be higher than the full magnetic
space group symmetry represented by the original magnetization density
$\vec{\mu}(\vec{r})$. This can occur even in systems that are usually
well described in terms of localized magnetic moments (see for example
our discussion of BiFeO$_3$ in Sec.~\ref{sec:BFO}). The localized
moment approach also neglects the possibility that the localized
current distribution around the atomic site, which gives rise to the
local magnetic dipole moment, simultaneously gives rise to a localized
toroidal dipole moment. Such ``atomic'' toroidal moments have been
discussed in the context of atomic multipole moments and can in
principle be measured by resonant x-ray
spectroscopy.\cite{DiMatteo/Joly/Natoli:2005} Here, we restrict our
discussion to the case of toroidal moments ``on the unit cell scale''
and disregard the possibility of toroidal contributions ``on the
atomic scale''.

Using Eq.~(\ref{t_spin}) we can straightforwardly evaluate the
toroidal moments of the arrangements shown in
Fig.~\ref{toroidal_cartoons}.  Taking the horizontal magnetic moments
to be spaced a distance $a$ apart along the $y$ direction, and the
vertical moments a distance $a$ apart along $x$, the toroidal moments
of arrangements a) and b) in Fig.~\ref{toroidal_cartoons} are $\vec{t}
= \pm as \hat{\vec{z}}$, where $s$ is the magnitude of the individual
magnetic moments.  The toroidal moment of
Fig.~\ref{toroidal_cartoons}c is $\vec{t} = -\frac{1}{2}as
\hat{\vec{z}}$, whereas it is zero for the arrangement shown in
Fig.~\ref{toroidal_cartoons}d, since in this case the moment vectors
are aligned parallel to the vector connecting the two sites.

\section{Origin dependence}
\label{origin}

\begin{figure}
\centerline{\includegraphics[width=0.95\columnwidth]{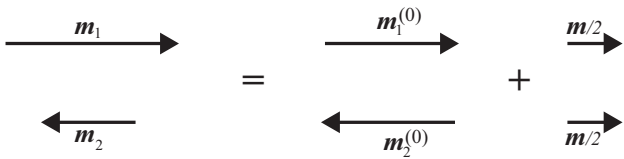}}
\caption{Decomposition of a ferrimagnetic arrangement of two localized
moments (left) into its fully compensated component (middle) and its
uncompensated ``ferromagnetic'' component (right). The ferromagnetic
component at each site is the total magnetic moment divided by the
total number of moments, $\tilde{\vec{m}}_\alpha = \frac{1}{2}(
\vec{m}_1 + \vec{m}_2 )$, and the compensated part is the difference
between the magnitude of the original local moment and the
uncompensated contribution.  }
\label{decompose_T_M}
\end{figure}

It can easily be seen that for systems with non-vanishing magnetic
dipole moment, 
\begin{equation}
\vec{m} = \frac{1}{2c} \int \vec{r} \times \vec{j}(\vec{r}) d^3r =
\int \vec{\mu}(\vec{r}) d^3r = \sum_\alpha \vec{m}_\alpha \quad,
\end{equation}
the toroidal moment in Eqs.~(\ref{t_formal}), (\ref{t_mu}), and
(\ref{t_spin}) depends on the choice of origin. In particular, for a
change of origin defined by $\vec{r} \rightarrow \vec{r}' = \vec{r} +
\vec{R}_0$ the toroidal moment changes as $\vec{t} \rightarrow
\vec{t}' = \vec{t} + 1/2 \vec{R}_0 \times \vec{m}$.

To further analyze this origin dependence, we decompose the original
magnetic moment distribution $\{\vec{m}_\alpha\}$ into two parts: a
totally compensated part $\{\vec{m}_\alpha^{(0)} = \vec{m}_\alpha -
\vec{m}/N \}$, with no net magnetization, $\sum_\alpha
\vec{m}_\alpha^{(0)} = 0$, and an uncompensated ``ferromagnetic'' part
$\{ \tilde{\vec{m}}_\alpha = \vec{m}/N \}$, where $\vec{m} =
\sum_\alpha \vec{m}_\alpha$ is the total magnetic moment and $N$ is
the total number of localized moments (see Fig.~\ref{decompose_T_M}
for a simple example). This results in a corresponding decomposition
of the toroidal moment into:
\begin{equation}
\label{t_comp}
\vec{t}^{(0)} = \frac{1}{2} \sum_{\alpha} \vec{r}_{\alpha} \times
\vec{m}_{\alpha}^{(0)} \quad ,
\end{equation}
and 
\begin{equation}
\tilde{\vec{t}} = \frac{1}{2} \, \bar{\vec{R}} \times \vec{m} \quad ,
\end{equation}
where $\bar{\vec{R}} = 1/N \sum_\alpha \vec{r}_\alpha$ is the average
magnetic moment position.  By construction, only the toroidal moment
$\tilde{\vec{t}}$ depends on the choice of origin, whereas the part
$\vec{t}^{(0)}$ is origin independent.

For a macroscopic toroidal moment to occur, the magnetic moment
distribution has to break \emph{both} time and space reversal
symmetries. In the compensated moment distribution $\{
\vec{m}^{(0)}_\alpha \}$ this can happen in several ways, depending
both on how the magnetic moments are \emph{oriented} and on how they
are \emph{positioned}. On the other hand the uncompensated
distribution $\{ \tilde{\vec{m}}_\alpha \}$ provides less freedom. Due
to the non-vanishing magnetic dipole moment, the configuration $\{
\tilde{\vec{m}}_\alpha \}$ always breaks time reversal
symmetry. However, the only possibility for such a ``ferromagnetic''
moment distribution to simultaneously break space inversion symmetry
is that the magnetic moments are \emph{positioned} in a
non-centrosymmetric way. A nonzero toroidal moment of the
uncompensated part of any moment distribution is therefore always
related to an inversion symmetry-breaking arrangement of the
underlying ionic lattice, whereas this does not necessarily have to be
the case for the compensated moment distribution $\{
\vec{m}_\alpha^{(0)} \}$, where the inversion symmetry can also be
lifted by the \emph{orientation} of the various magnetic
moments.\cite{footnote:compensated}

It will become clear in the following section, that only differences
in toroidal moment should have any physical significance. In the case
of the origin-dependent part $\tilde{\vec{t}}$, such differences in
toroidal moment must be related to corresponding displacements of the
moment positions, and the change in the toroidal moment is then given
by $\Delta \tilde{\vec{t}} = 1/2 \Delta \bar{\vec{R}} \times \vec{m}$,
with $\Delta \bar{\vec{R}} = 1/N \sum_\alpha \Delta \vec{r}_{\alpha}$
and $\Delta \vec{r}_{\alpha}$ being the displacements of the
individual magnetic moments. Thus, if a consistent choice of origin is
used for the initial and final configuration, the corresponding change
in the toroidal moment is a well-defined physical quantity. For
example, if the initial reference configuration is centrosymmetric,
then the change in toroidal moment resulting from a symmetry-breaking
structural distortion can be interpreted as the spontaneous toroidal
moment of the system.

Earlier applications of Eq.~(\ref{t_spin}) did not perform the
explicit decomposition described above, but instead evaluated the
toroidal moment with respect to the ``center of the unit
cell'',\cite{Popov_et_al:1998} without specifying exactly how this
center of the unit cell is defined. We point out that the origin
dependent contribution to the toroidal moment $\tilde{\vec{t}}$
vanishes, if the ``center of the unit cell'' defined by
$\bar{\vec{R}}$ is taken as the origin.\cite{footnote:rbar} However,
it is important to realize that for cases where the toroidal moment
changes as a result of a structural distortion, $\bar{\vec{R}}$ in
general also changes (see the discussion in the previous
paragraph). In such cases the origin should be taken to be the same
for both structural modifications.

Finally, we note that in the case of a non-localized magnetization
density, the ``uncompensated'' part of $\vec{\mu}(\vec{r})$ would
correspond to a uniform, i.e. perfectly homogeneous
($\vec{r}$-independent), magnetization density $\tilde{\vec{\mu}} =
\vec{m}/V$, where $V$ is the total volume of the system. Since such a
perfectly homogeneous magnetization density $\tilde{\vec{\mu}}$ can
never break space-reversal symmetry, it does not contribute at all to
the toroidal moment of the system. Thus, in a non-localized
description, the decomposition into compensated and uncompensated
parts results in a perfect separation between dipolar and toroidal
contributions to the magnetization density. Due to the
``inhomogeneity'' that is intrinsic to the localized moment
description, the decomposition is not fully complete in this case, and
the toroidal contribution of the ``ferromagnetic'' part has to be
analyzed separately.

In our explicit examples in Secs.~\ref{chain} and \ref{examples} we
will only consider systems with fully compensated moment
configurations and postpone the further analysis of toroidal moments
resulting from ``ferromagnetic'' configurations to future work.

\section{The case of a periodic bulk system}
\label{sec:bulk}

We now proceed to the case of bulk periodic crystals with an infinite
periodic arrangement of magnetic moments. We begin by defining the
toroidal moment per unit volume or \emph{toroidization} $\vec{T} =
\vec{t}/V$, where $V$ is the volume of the system with toroidal moment
$\vec{t}$.  Then, for a large finite system containing $N$ identical
unit cells each of volume $\Omega$:
\begin{align}
  \vec{T} & = \frac{1}{2N \Omega} \sum_{\alpha}
  \vec{r}_{\alpha} \times \vec{m}_{\alpha} \\ & = \frac{1}{2N \Omega}
  \sum_{n,i} (\vec{r}_i+ \vec{R}_n) \times \vec{m}_i \quad .
\end{align}
Here, $\vec{r}_i$ are the positions of the magnetic moments
$\vec{m}_i$ relative to the same (arbitrary) point within each unit
cell, $\vec{R}_n$ is a ``lattice vector'' with index $n$, and we have
used the fact that the orientation of the magnetic moments is the same
in each unit cell. The summation over $i$ indicates the summation over
all moments within a unit cell, and that over $n$ indicates the
summation over all unit cells.  Expanding the cross product, we
obtain:
\begin{align}
\vec{T} & = \frac{1}{2\Omega} \sum_i \vec{r}_i \times \vec{m}_i +
\frac{1}{2N \Omega} \sum_{n} \vec{R}_n \times \sum_i \vec{m}_i \\ 
& =  \frac{1}{2\Omega} \sum_i \vec{r}_i \times \vec{m}_i +
\frac{1}{2N^2 \Omega} \sum_{n} \vec{R}_n \times \vec{m} \\ 
& = \frac{1}{2\Omega} \sum_i \vec{r}_i \times \vec{m}_i \quad ,
\end{align}
where in the last step we have assumed that the sum over all lattice
vectors contains both $\vec{R}_n$ and $-\vec{R}_n$, so that $\sum_n
\vec{R}_n = 0$. This is true for any infinite Bravais lattice. Thus,
the toroidal moment of a system of $N$ unit cells is just $N$ times
the toroidal moment evaluated for one unit cell, and the corresponding
toroidizations are identical.

In an infinite periodic solid, we have a freedom in choosing the basis
corresponding to the primitive unit cell of the crystal. In
particular, we can translate any spin of the basis by a lattice vector
$\vec{R}_n$ without changing the overall periodic
arrangement. However, such a translation of magnetic moment
$\vec{m}_i$ by $\vec{R}_n$ leads to a change in the toroidization as
follows:
\begin{equation}
\label{quantum}
\Delta \vec{T}_{ni} = \frac{1}{2\Omega} \vec{R}_n \times \vec{m}_i \quad .
\end{equation}
The freedom in choosing the basis corresponding to the primitive unit
cell thus leads to a multivaluedness of the toroidization with respect
to certain ``increments'' (defined by Eq.~({\ref{quantum})) for each
magnetic sub-lattice $i$ and lattice vector $\vec{R}_n$.

This multivaluedness of the toroidization is in strong analogy to the
modern theory of electric
polarization,\cite{King-Smith/Vanderbilt:1993,Vanderbilt/King-Smith:1993,Resta:1994}
where the polarization changes by $e\vec{R}_n/\Omega$ if one
translates an elementary charge $e$ by a multiple of a lattice vector
$\vec{R}_n$. The resulting multivaluedness has led to the concept of
the ``polarization lattice'' corresponding to a bulk periodic
solid,\cite{Vanderbilt/King-Smith:1993} with $e\vec{R}_n/\Omega$
called the ``polarization quantum'' if $\vec{R}_n$ is one of the three
primitive lattice vectors.  Eq.~(\ref{quantum}) suggests the existence
of an analogous ``toroidization lattice''. However, the vector product
in Eq.~(\ref{quantum}) and our classical treatment of magnetic moments
can lead to arbitrary projections of $\vec{m}_i$ on a certain axis,
and therefore the structure of the ``toroidization lattice'' is quite
different from that of the polarization lattice. In the most general
case, if there are $r$ magnetic basis atoms in the primitive unit
cell, there can be 3$r$ linearly independent toroidization
increments. This can lead to cases, where multiple incommensurate
increments exist along certain crystallographic directions. In
addition, for a collinear moment configuration no toroidization
component is allowed parallel to the global magnetic axis. Thus, the
set of allowed toroidization values does not necessarily have the same
translation symmetry as the corresponding crystal structure and does
not necessarily form a Bravais lattice, whereas this is always true in
the case of the electric polarization. In practice, the magnetic
symmetry of the system can significantly reduce the number of linearly
independent toroidization increments (see Sec.~\ref{examples} for some
realistic examples.

In spite of the difficulties associated with the multivaluedness of
the polarization, it is now recognized that only differences in the
polarization lattices between different ionic configurations are in
fact measurable quantities, such as for example the difference between
two oppositely polarized states of a ferroelectric crystal or between
a centrosymmetric non-polar reference structure and the actual polar
crystal. These differences are the same for each point of the
polarization lattice and are thus well-defined quantities.  In analogy
with the case of the electric polarization we suggest that only
differences in the set of allowed toroidization values, corresponding
to two different bulk configurations, are physically observable
quantities, such as the difference in toroidizations between two
domain states of a ferrotoroidic, or the difference between a
ferrotoroidic state and its non-toroidic paraphase.  Such quantities
can be obtained by monitoring the change in toroidal moment on one
arbitrarily chosen \emph{branch} within the allowed set of values,
when transforming the system from the initial to the final state along
a well-defined path.

Finally, we emphasize that the multivaluedness of the toroidization
and the possible origin dependence of the toroidization are two
independent features with different origins. Both features are rooted
in the fundamental definition of the toroidal moment in terms of the
position operator $\vec{r}$, but whereas the origin dependence appears
in both finite and infinite systems if there is a non-vanishing
magnetic dipole moment, the multivaluedness is caused by the periodic
boundary conditions in a bulk solid, and is independent of an
eventually non-vanishing magnetic dipole moment.

\section{A one-dimensional example}
\label{chain}

\subsection{The periodic non-toroidal state}
\label{afm_chain_a}

\begin{figure}
\centerline{\includegraphics[width=0.8\columnwidth]{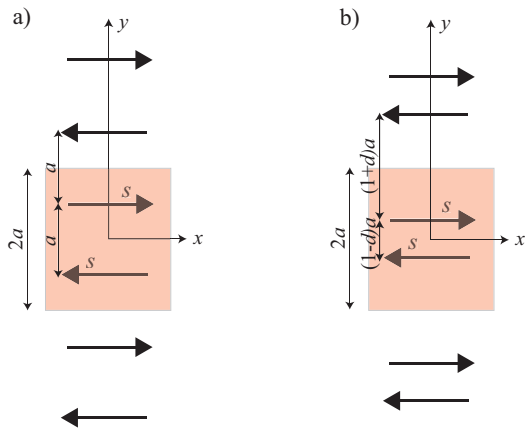}}
\caption{Calculation of the toroidization for two different
  one-dimensional antiferromagnetic periodic arrangements of magnetic
  moments. Our choice of unit cell is indicated by the shaded area in
  each case. a) shows a non-toroidal state, which is space-inversion
  symmetric with respect to each moment site. b) is a toroidal
  state.}
\label{lattice}
\end{figure}

To illustrate some consequences of the multivaluedness of the
toroidization in periodic systems described in the previous section,
we now consider the example of a one-dimensional antiferromagnetic
chain of equally spaced magnetic moments as shown in
Fig.~\ref{lattice}a. The moments, with magnitude $s$, are spaced a
distance $a$ apart from each other along the $y$ axis, and are
alternating in orientation along $\pm x$. Thus, the unit cell length
is $2a$ and there are two oppositely oriented magnetic moments in each
unit cell. Since this configuration does not possess a macroscopic
magnetic dipole moment, the corresponding toroidal moment is
origin-independent, and a decomposition into compensated and
uncompensated parts is not required.

The arrangement of magnetic moments in Fig.~\ref{lattice}a is
space-inversion symmetric with respect to each moment site and thus
cannot exhibit a macroscopic toroidal moment. Furthermore, even though
the arrangement in Fig.~\ref{lattice}a breaks time reversal symmetry
\emph{microscopically}, there exists a symmetry transformation which
combines time inversion with a translation of all moments by the
distance $a$ along the $y$ direction. According to Neumann's principle
(see e.g. Ref.~\onlinecite{Birss:book}), the macroscopic properties of
a system cannot depend on such microscopic translations, i.e the
macroscopic properties are determined by the \emph{point group} of the
system and not by its \emph{space group}. Therefore, time reversal
symmetry is not broken \emph{macroscopically} for the moment
configuration in Fig.~\ref{lattice}a and its point group contains time
inversion as a symmetry element. No macroscopic toroidal moment can
thus result from this configuration, in spite of the fact that an
isolated unit cell would exhibit a toroidal moment.

The toroidal moment of the single unit cell highlighted in
Fig.~\ref{lattice}a, calculated using Eq.~(\ref{t_spin}), is identical
to that calculated for the finite spin configuration in
Fig.~\ref{toroidal_cartoons}c, i.e. $\vec{t} = -\frac{1}{2}sa
\hat{\vec{z}}$, and the corresponding toroidization, $\vec{T} =
\vec{t}/\Omega = -\frac{s}{4} \hat{\vec{z}}$ (since the ``volume''
$\Omega$ of the one-dimensional unit cell is just its length, $2a$).
The elementary toroidization increment in this case is $\Delta\vec{T}
= \pm \frac{s}{2} \hat{\vec{z}}$, which means that the toroidization
of the unit cell is exactly equal to one half of the toroidization
increment, and the allowed toroidization values for the periodic
arrangement are $\vec{T}_n = (\frac{1}{2} + n)\frac{s}{2}
\hat{\vec{z}}$, where $n$ can be any integer number.

We see that in our example the allowed toroidization values form a
one-dimensional lattice of values, centrosymmetric around the
origin. This is analogous to the case of the electric polarization,
where the polarization lattice is invariant under all symmetry
transformations of the underlying crystal structure. In particular,
the polarization lattice corresponding to a centrosymmetric crystal
structure has to be inversion symmetric. This can be achieved either
by a lattice that includes the point $\vec{P} = 0$, or by the same
lattice but shifted from the origin by exactly half a polarization
quantum. We see that the same holds true for the toroidization of our
one-dimensional example, so that a centrosymmetric set of
toroidization values can be understood as representing a non-toroidal
state of the corresponding system.

We emphasize again, however, that the toroidization case is slightly more 
involved than the case of the electric polarization. In particular, due to 
the fact that several incommensurate toroidization increments can exist 
along the same crystallographic direction, there are many more allowed 
values of the toroidization in the non-toroidal state than for the polarization
in the non-polar state, as can be seen for the
system BaNiF$_4$ discussed in Sec.~\ref{sec:bnf}.

\subsection{Toroidal state and changes in toroidization}

In order to obtain a nontrivial ``macroscopic'' toroidization the
system has to break both space and time inversion symmetry. In the
case of the one-dimensional antiferromagnetic chain this can be
achieved by ``spin pairing'', i.e.  if the distances between
neighboring magnetic moments alternate as shown in
Fig.~\ref{lattice}b.  Here the magnetic moments of magnitude $s$ are
spaced alternately a distance of $(1-d)a$ and $(1+d)a$ apart from each
other along the $y$ axis ($-1 < d < 1$), and again are alternating in
orientation along $\pm x$. The non-toroidal example above corresponds
to $d = 0$.  Since the unit cell size is the same as in the
non-toroidal case, the elementary toroidization increment is again
$\Delta\vec{T} = \pm \frac{s}{2} \hat{\vec{z}}$. The toroidization of
the unit cell indicated in Fig.~\ref{lattice}b is $\vec{T} = -
(1-d)\frac{s}{4} \hat{\vec{z}}$, so that the allowed values of
$\vec{T}$ for the full periodic arrangement are:
\begin{equation}
\label{T-chain}
\vec{T}_n(d) = \left( n - \frac{1-d}{2} \right) \frac{s}{2} \hat{\vec{z}}
\quad .
\end{equation}
Fig.~\ref{t_of_d} shows the allowed toroidization values as a function
of the displacement $d$ of the spins from their positions in the
centrosymmetric, non-toroidal state.

\begin{figure}
\centerline{\includegraphics[width=0.9\columnwidth]{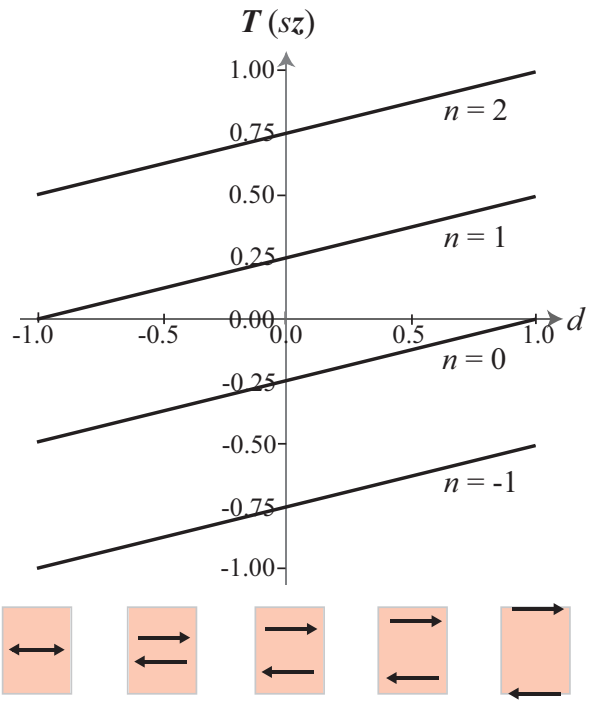}}
\caption{Allowed values of the toroidization for the antiferromagnetic
  chain of Fig.~\ref{lattice} as a function of displacement $d$ from
  the non-toroidal case ($d=0$). The cartoons at the bottom indicate
  the corresponding positions of the magnetic moments within the unit
  cell. }
\label{t_of_d}
\end{figure}

The change in toroidization between two configurations with $d=d_1$
and $d=d_2$ for a certain branch $n$ is given by:
\begin{equation}
\vec{T}_n(d_2) - \vec{T}_n(d_1) = \frac{s}{4} \left( d_2 - d_1 \right)
\hat{\vec{z}} \quad ,
\label{T_model}
\end{equation}
i.e. it is independent of the branch index $n$. In particular, if the
non-centrosymmetric distortion is inverted ($d_1=d_0$, $d_2=-d_0)$, the
change in toroidization is $2T_\text{s} = \frac{sd_0}{2}$ so that
$T_\text{s} = \frac{sd_0}{4}$ can be interpreted as the
\emph{spontaneous toroidization}, again in analogy to the case of the
electric polarization, where the spontaneous polarization is given by
the branch-independent change in polarization compared to a
centrosymmetric reference structure.

Another possible way to alter the toroidization is by changing the
orientation of the magnetic moments instead of changing their
positions. In particular, we expect that a full 180$^\circ$ rotation
of all magnetic moments, which is equivalent to the operation of time
reversal, should invert the macroscopic ``spontaneous toroidization'',
and should therefore lead to the same change $2T_\text{s}$ as
discussed above. If we allow the magnetic moments to rotate out of the
$x$ direction, while preserving the antiparallel alignment of the two
basis moments, the toroidization along the $z$ direction is given by
\begin{equation}
T^z_n(d,\alpha) = \left( n - \frac{1-d}{2} \right) \frac{s}{2}
\cos\alpha \quad ,
\end{equation}
where $\alpha$ is the angle between the magnetic moments and the $x$
direction. The change in toroidization for a full 180$^\circ$ rotation
of the moments is thus:
\begin{equation}
\label{t-change}
T^z_n(d_0,180^\circ) - T^z_n(d_0,0^\circ) = -\frac{sd_0}{2} + s(2n+1)
\quad ,
\end{equation}
and apparently depends on the branch index $n$. However, if one
calculates the same change in toroidization for the non-toroidal state
with $d=0$, one obtains:
\begin{equation}
\label{improper}
T^z_n(0,180^\circ) - T^z_n(0,0^\circ) = s(2n+1) \quad .  
\end{equation}
Obviously, in this case the corresponding change in macroscopic
toroidization should be zero, since both the initial and final states
(and all intermediate states) correspond to a non-toroidal
configuration and thus $T_\text{s}=0$. If one subtracts the \emph{improper}
change in $T^z$, Eq.~(\ref{improper}), from the change in
toroidization calculated in Eq.~(\ref{t-change}), one obtains the
\emph{proper} change in toroidization $2T_\text{s} = \frac{sd_0}{2}$,
which is identical to one obtained by inverting the non-centrosymmetric
distortion $d$. Here, we use the terminology ``proper'' and
``improper'' in analogy to the case of the proper and improper
piezoelectric response, \cite{Vanderbilt:2000} where a similar branch
dependence is caused by volume changes of the unit cell, and the
improper piezoelectric response has to be subtracted appropriately.

\begin{figure}
\centerline{\includegraphics[width=0.95\columnwidth]{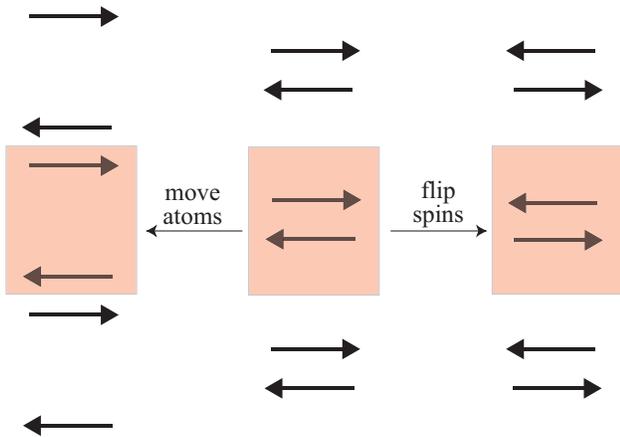}}
\caption{Effect on the magnetic moment configuration of
Fig.~\ref{lattice}b (center) of a reversal of all magnetic moments
(right) and of a reversal of the non-centrosymmetric distortion $d$
(left). Note that the right and left final states are identical, with
the moments on the left translated by half a unit cell compared to
those on the right.}
\label{spin_flip}
\end{figure}

Fig.~\ref{spin_flip} shows the initial and final states for the two
cases where either the moment displacements or the magnetic moment
directions are inverted. The two final states are equivalent except
for a translation of all moments by half a unit cell along $y$, which,
again due to Neumann's principle, is irrelevant for the macroscopic
properties. The spontaneous toroidization of the state on the left in
Fig.~\ref{spin_flip} is therefore the same as for the state on the
right side of Fig.~\ref{spin_flip}.

\section{Toroidizations for some example materials}
\label{examples}

To further illustrate the concept of toroidal moments in crystalline
solids and to investigate the consequences of the definitions and
simplifications outlined in the preceding sections for real systems,
we now evaluate the toroidizations of four example materials:
BaNiF$_4$, LiCoPO$_4$, GaFeO$_3$, and BiFeO$_3$. All these materials
have been discussed recently in the context of multiferroics,
magneto-electric coupling, or ferrotoroidics. BaNiF$_4$ and BiFeO$_3$
are both antiferromagnetic ferroelectrics, with additional weak
ferromagnetism in the case of BiFeO$_3$, and the possible coupling
between the various order parameters in these systems has recently
been studied using first principles
techniques.\cite{Ederer/Spaldin:2006,Ederer/Spaldin:2005} LiCoPO$_4$,
which is not ferroelectric, exhibits a rather large linear
magneto-electric effect,\cite{Rivera:1994} and the observation of
ferrotoroidic domains in this material using nonlinear optical
techniques has been reported.\cite{VanAken_et_al:2007} GaFeO$_3$ is a
magnetic piezoelectric which exhibits a strong asymmetry in the
magneto-electric tensor. This asymmetry has been interpreted to result
from a non-vanishing toroidal moment in this
material.\cite{Popov_et_al:1998}

\subsection{BaNiF$_4$}
\label{sec:bnf}

BaNiF$_4$ belongs to an isostructural family of antiferromagnetic
ferroelectrics with composition Ba$M$F$_4$, where $M$ can be Mn, Fe,
Co, or Ni.\cite{Schnering/Bleckmann:1968,Eibschuetz_et_al:1969} It was
recently shown that BaNiF$_4$ exhibits two distinct antiferromagnetic
order parameters and that the order parameter corresponding to the
``weak'' (secondary) antiferromagnetic order can be reversed by using
an electric field.\cite{Ederer/Spaldin:2005} The crystallographic
structure of BaNiF$_4$ is orthorhombic corresponding to space group
$Cmc2_1$.\cite{Schnering/Bleckmann:1968,Eibschuetz_et_al:1969} The
magnetic space group is $P_a2_1$,\cite{Cox_et_al:1970} which contains
a non-primitive translation along the $a$ direction combined with time
inversion.\cite{Bradley/Cracknell:Book} This reflects the fact that
the antiferromagnetic ordering in BaNiF$_4$ leads to a unit cell
doubling compared to the paramagnetic phase. The corresponding
(macroscopic) magnetic point group is thus $21'$, which does not break
time inversion, similar to the case of the one-dimensional
antiferromagnetic chain discussed in
Sec.~\ref{afm_chain_a}. Therefore, neither a macroscopic magnetization
nor a macroscopic toroidal moment is allowed in this symmetry. We note
that, in general, when the magnetic ordering leads to a unit cell
doubling compared to the paramagnetic phase, then the system is always
macroscopically time reversal symmetric and therefore non-toroidal.

\begin{table}
\caption{Positions $\vec{r}_i$ and moment directions $\vec{m}_i$ of
the magnetic Ni cations in BaNiF$_4$. $a$, $b$ and $c$ are the
orthorhombic lattice constants, $\delta$ represents an internal
structural parameter, $\theta$ is the angle between the magnetic
moments and the orthorhombic $b$ axis, and $S$ is the magnitude of the
Ni magnetic moment. The corresponding lattice vectors are: $\vec{a}_1
= (a,b,0)$, $\vec{a}_2 = (a/2,-b/2,0)$, and $\vec{a}_3 = (0,0,c)$.}
\label{bnf-struc}
\begin{ruledtabular}
\begin{tabular}{l|ccc|ccc}
site & $r^x_i/a$ & $r^y_i/b$ & $r^z_i/c$ & $m^x_i$ & $m^y_i$ & $m^z_i$ \\ 
\hline
Ni & 0 &  $\delta$ & 0              & 0 & $S \cos\theta$  & $S \sin\theta$ \\
Ni & 0 & $-\delta$ & $\tfrac{1}{2}$ & 0 & $-S \cos\theta$ & $S \sin\theta$ \\
Ni & 1 &  $\delta$ & 0              & 0 & $-S \cos\theta$ & $-S \sin\theta$ \\
Ni & 1 & $-\delta$ & $\tfrac{1}{2}$ & 0 & $S \cos\theta$  & $-S \sin\theta$
\end{tabular}
\end{ruledtabular}
\end{table}

Nevertheless it is instructive to examine the effect of the periodic
boundary conditions and the resulting multivaluedness of the
toroidization for this trivial case. The magnetic unit cell of
BaNiF$_4$ contains 4 magnetic Ni ions, whose positions and spin
directions are listed in Table~\ref{bnf-struc}. A decomposition into
fully compensated and uncompensated components is not necessary, since
there is no macroscopic magnetization in this system. Application of
Eqs.~(\ref{t_spin}) and (\ref{quantum}) leads to a ``toroidization
lattice'' of the form:
\begin{multline}
\label{t_bnf}
\vec{T}_{klmn} = \frac{S}{2 \Omega} \left\{ \cos\theta \left( \begin{array}{c} 
c k \\ 
0 \\
\frac{a}{2} (l-m+2n) 
\end{array} \right) \right. \\ \left. 
+ \sin\theta \left( \begin{array}{c} 
\frac{b}{2} (l+m-2n) \\ 
\frac{b}{2} (l+m+2n+4) \\
0
\end{array} \right) \right\} \quad ,
\end{multline}
where $\theta$ is the canting angle of the magnetic moments (of
magnitude $S$) relative to the orthorhombic $b$ direction, and $a$,
$b$, $c$ are the orthorhombic lattice parameters.  $k$, $l$, $m$, and
$n$ are arbitrary integer numbers corresponding to four different
independent toroidization increments in this system. Due to the
pairwise collinear spin structure in BaNiF$_4$ the original $4 \times
3 = 12$ increments according to Eq.~(\ref{quantum}) are reduced to $2
\times 3 = 6$. Additional symmetries reduce the number of independent
increments to 4. Due to the base-centered orthorhombic Bravais lattice
of BaNiF$_4$, these toroidization increments are in general neither
parallel to the cartesian coordinate directions nor perpendicular to
each other. It can be seen from Eq.~(\ref{t_bnf}) that for a general
value of the angle $\theta$, the allowed values of $\vec{T}$ do not
form a Bravais lattice, and the multivaluedness of $\vec{T}$ is more
complex than in the case of the electric polarization. Nevertheless,
the set of allowed toroidization values in BaNiF$_4$ is inversion
symmetric as required by the time-symmetric point group.  In the
absence of canting ($\theta = 0^\circ$), the toroidization and the
toroidization increment along the $b$ direction would be zero; however
we see that for small $\theta$ there is a small non-zero increment of
the toroidization along this direction, reflecting the small component
of the magnetic moments perpendicular to $b$.  Thus, the example of
BaNiF$_4$ shows that the structure of the set of allowed toroidization
values is in general more complex than the ``polarization lattice'' in
the modern theory of electric polarization in crystalline solids.

\subsection{LiCoPO$_4$}

The observation of ferrotoroidic domains in LiCoPO$_4$ using nonlinear
optical techniques has been reported in
Ref.~\onlinecite{VanAken_et_al:2007}. LiCoPO$_4$ crystallizes in the
olivine structure with the orthorhombic space group
$Pnma$,\cite{Newnham/Redman:1965} and originally it was believed that
the magnetic moments of the four Co ions in the unit cell are
antiferromagnetically aligned along the orthorhombic $b$
direction.\cite{Santoro/Segal/Newnham:1966} This moment configuration
corresponds to the magnetic space group $Pnma'$, which contains the
operation of simultaneous time and space inversion, but not the pure
time and space inversion separately, and thus allows the existence of
a macroscopic toroidal moment. Furthermore, the toroidal moment is
required to be aligned parallel to the $c$ axis. Recently, it was
found that the magnetic moments in LiCoPO$_4$ are rotated slightly
away from the $b$ direction by an angle $\theta \approx 4.6^\circ$,
while preserving the overall collinear magnetic
structure.\cite{Vaknin_et_al:2002} For a moment rotation within the
$b$-$c$ plane, the rotated moment configuration corresponds to a lower
symmetry with the magnetic point group $2'/m$, which allows a toroidal
moment also along the orthorhombic $b$ direction. At the moment, it is
not clear what the primary order parameter for this additional
symmetry lowering is, although since it does not change the relative
antiferromagnetic arrangement of the spins a toroidal origin has been
proposed.\cite{VanAken_et_al:2007} Furthermore, a weak magnetization
has been measured along the $b$ direction,\cite{Kharchenko_et_al:2001}
which indicates an even lower point group symmetry of $2'$. Since the
small magnetization occurs along the $b$ direction, i.e. parallel to
the direction of the antiferromagnetic alignment, it has been
described as weak ferrimagnetism.

\begin{table}
\caption{Positions $\vec{r}_i$ and magnetic moment directions
$\vec{m}_i$ of the Co cations in LiCoPO$_4$, according to
Ref.~\onlinecite{Vaknin_et_al:2002} and assuming a moment rotation
towards the $c$ direction. $\epsilon$ and $\delta$ are internal
structural parameters corresponding to Wyckoff positions 4$c$ of the
$Pnma$ space group. $\theta$ is the angle between the magnetic moments
and the $b$ axis and $S$ is the magnitude of the Co magnetic
moment. Experimental values are $\epsilon=0.0286$, $\delta=0.0207$,
and $\theta \approx 4.6^\circ$.\cite{Vaknin_et_al:2002} The
orthorhombic lattice parameters are $a=10.20$\,\AA, $b=5.92$\,\AA, and
$c=4.70$\,\AA.\cite{Newnham/Redman:1965}}
\label{lnpo-struc}
\begin{ruledtabular}
\begin{tabular}{l|ccc|ccc}
site & $r^x_i/a$ & $r^y_i/b$ & $r^z_i/c$ & $m^x_i$ & $m^y_i$ & $m^z_i$ \\ 
\hline
Co & $1/4+\epsilon$ & $1/4$ & $-\delta$ & 0 & $-S \cos\theta$ & $-S \sin\theta$ \\
Co & $1/4-\epsilon$ & $-1/4$ & $1/2-\delta$ & 0 & $S \cos\theta$ & $S \sin\theta$ \\
Co & $-1/4-\epsilon$ & $-1/4$ & $\delta$ & 0 & $S \cos\theta$ & $S \sin\theta$ \\
Co & $-1/4+\epsilon$ & $1/4$ & $1/2+\delta$ & 0 & $-S \cos\theta$ & $-S \sin\theta$
\end{tabular}
\end{ruledtabular}
\end{table}

Here we calculate the toroidization corresponding to the fully
compensated antiferromagnetic configuration with magnetic point group
$2'/m$, where all spins are rotated from the $b$ direction towards the
$c$ direction by an angle $\theta$. The corresponding positions and
spin directions of the four magnetic Co cations within the unit cell
are listed in Table~\ref{lnpo-struc}. The toroidization resulting from
Eqs.~(\ref{t_spin}) and (\ref{quantum}) is:
\begin{equation}
\vec{T}_{nml} =  \frac{S}{2\Omega} \left\{ 
\cos\theta \left( \begin{array}{c} 
c m \\ 
0 \\
(4\epsilon \mp l) a 
\end{array}  \right) + 
\sin\theta \left(  \begin{array}{c}
b n \\ 
(4\epsilon \pm l )  a \\ 
0 
\end{array} \right) \right\} \quad .
\end{equation}
Here, $a$, $b$ and $c$ are the orthorhombic lattice constants and $n$,
$m$, and $l$ are arbitrary integer numbers. It can be seen that there
is only a trivial component of $\vec{T}$ along the $a$ direction and
that there is a nontrivial part of the toroidal moment proportional to
$\epsilon$, which rotates from the $c$ direction for $\theta=0^\circ$
towards the $b$ direction for $\theta=90^\circ$. For $\epsilon=0$ the
system is centrosymmetric and thus non-toroidal. The spontaneous
toroidization is given by $T_s = \frac{2S\epsilon a}{\Omega}$. For the
experimentally reported parameters listed in Table~\ref{lnpo-struc}
and using the formal magnetic moment $S=3\mu_\text{B}$ of Co$^{2+}$,
the corresponding value is $\vec{T}_s = 6.17 \cdot 10^{-3}
\mu_\text{B}/\textrm{\AA}^2 ( \hat{\vec{z}} \cos\theta + \hat{\vec{y}}
\sin\theta )$, corresponding to a spontaneous toroidal moment per unit
cell of $\vec{t}_\text{s} = 1.75 \mu_\text{B} \textrm{\AA}\,(
\hat{\vec{z}} \cos\theta + \hat{\vec{y}} \sin\theta )$. Thus, the
magnetic moment rotation described by $\theta$ results in a
corresponding rigid rotation of the toroidization.

\subsection{GaFeO$_3$}

Ga$_{2-x}$Fe$_x$O$_3$ was the first material that was found to exhibit
both piezoelectricity and a macroscopic
magnetization.\cite{Remeika:1960} It was also the first known material
with a spontaneous magnetization that simultaneously exhibits a linear
magneto-electric effect.\cite{Rado:1964} The system
Ga$_{2-x}$Fe$_x$O$_3$ has been studied recently because of its
interesting optical properties, which result from the simultaneous
breaking of both space and time inversion symmetries in this
material.\cite{Kubota_et_al:2004,Jung_et_al:2004} In addition, an
asymmetry of the magneto-electric tensor has been measured and was
attributed to the existence of a toroidal moment in this
system.\cite{Popov_et_al:1998}

The crystal structure (space group $Pc2_1n$) of Ga$_{2-x}$Fe$_x$O$_3$
contains four inequivalent cation sites, one with tetrahedral oxygen
coordination and three different octahedrally coordinated
sites.\cite{Abrahams/Reddy/Bernstein:1965} The Fe cations occupy
predominantly two of the octahedral sites (called Fe1 and Fe2), but
there is also a sizable Fe occupation on the third octahedral site
(Ga2), whereas the tetrahedral site (Ga1) is occupied mainly by
Ga. The exact occupation of the various cation sites depends on the
composition $x$ as well as on the preparation technique and sample
history.\cite{Arima_et_al:2004}

Two different magnetic structures have been proposed for this system,
both of which are consistent with the magnetic space group $Pc'2_1'n$,
which allows for a macroscopic magnetization along the
crystallographic $c$ direction. Abrahams and Reddy originally proposed
a canted antiferromagnetic structure with compensating magnetic
moments within the $a$-$b$ plane and a net magnetization along
$c$.\cite{Abrahams/Reddy:1964} On the other hand Arima {\it et al.}
(Ref.~\onlinecite{Arima_et_al:2004}) recently interpreted their data
in terms of a collinear ferrimagnetic configuration suggested in
Ref.~\onlinecite{Delapalme:1967}, where all spins are oriented either
parallel or antiparallel to the $c$ direction. In both cases, the
relative orientation of the four symmetry-related magnetic moments on
the Fe1 sites or the Fe2 sites, respectively, is dictated by the
magnetic space group $Pc'2_1'n$.

We note that in the magnetic configuration used by Arima {\it et al.}
(Ref.~\onlinecite{Arima_et_al:2004}) the net magnetization stems
mainly from the intersite disorder; the system is a perfectly
compensated antiferromagnet if (i) all Fe1 and Fe2 sites are occupied
by Fe cations, (ii) there is no Fe occupation of the Ga1 and Ga2
sites, and (iii) the magnetic moments on the two Fe sites are the
same. Here, we consider only this perfectly compensated configuration
with no site disorder and composition $x=1$, i.e. all Fe1 and Fe2
sites are occupied by Fe$^{3+}$ cations and all Ga1 and Ga2 sites are
occupied by Ga$^{3+}$ cations. Thus, the magnetic configuration
discussed in the following does not have a net magnetization. We point
out that in general the magnetic and toroidal properties will depend
on the exact occupation numbers of the various cation sites. The
positions of the Fe sites as well as the corresponding moment
directions are listed in Table~\ref{struc:GFO}.

\begin{table}
\caption{Positions $\vec{r}_i$ and magnetic moment directions
$\vec{m}_i$ of the Fe sites in GaFeO$_3$ with magnetic space group
$Pc'2_1'n$. Both Fe sites correspond to Wyckoff positions 4$a$. The
magnetic configuration is that discussed in
Ref.~\onlinecite{Arima_et_al:2004}. $S$ is the magnitude of the Fe
magnetic moment, which is assumed to be identical on both sites and
$a$, $b$, $c$ are the usual orthorhombic lattice parameters.}
\label{struc:GFO}
\begin{ruledtabular}
\begin{tabular}{l|ccc|ccc}
site & $r^x/a$ & $r^y/b$ & $r^z/c$ & $m_i^x$ & $m_i^y$ & $m_i^z$ \\ 
\hline
Fe1 & $x_1$              & $y_1$              & $z_1$              & 0 & 0 & $S$ \\
Fe1 & $\tfrac{1}{2}-x_1$ & $y_1$              & $\tfrac{1}{2}+z_1$ & 0 & 0 & $S$ \\
Fe1 & $\tfrac{1}{2}+x_1$ & $y_1-\tfrac{1}{2}$ & $\tfrac{1}{2}-z_1$ & 0 & 0 & $S$ \\
Fe1 & $1-x_1$            & $y_1-\tfrac{1}{2}$ & $1-z_1$            & 0 & 0 & $S$ \\
Fe2 & $x_2$              & $y_2$              & $z_2$              & 0 & 0 & $-S$ \\
Fe2 & $\tfrac{1}{2}-x_2$ & $y_2$              & $\tfrac{1}{2}+z_2$ & 0 & 0 & $-S$ \\
Fe2 & $\tfrac{1}{2}+x_2$ & $y_2-\tfrac{1}{2}$ & $\tfrac{1}{2}-z_2$ & 0 & 0 & $-S$ \\
Fe2 & $1-x_2$            & $y_2-\tfrac{1}{2}$ & $1-z_2$            & 0 & 0 & $-S$ \\
\end{tabular}
\end{ruledtabular}
\end{table}

The magnetic space group $Pc'2_1'n$ breaks both space and time
reversal symmetries and thus allows the existence of a macroscopic
toroidal moment. Evaluation of Eqs.~(\ref{t_spin}) and (\ref{quantum})
for the positions and moment directions listed in
Table~\ref{struc:GFO} leads to the toroidization:
\begin{equation}
\label{GFO:t2}
\vec{T}_{nm} = \frac{S}{2\Omega} \left(  \begin{array}{c}
    4(y_1 - y_2)b + nb \\
    ma \\
    0 \end{array}  \right) \quad .
\end{equation}
It can be seen that there is a nontrivial toroidization along the $a$
direction, as dictated by the magnetic space group symmetry, whereas
the component along the $b$ direction represents only the trivial
increment resulting from the periodic boundary conditions, and the
component along $c$ is zero. The macroscopic toroidization along $a$
depends on the difference of the coordinates $y_1$ and $y_2$ of the
two different Fe sites along $b$.

One can verify that for $y_1 - y_2 = \frac{1}{4}l$ (for any integer
$l$) the ``magnetic lattice'', i.e. the spatial arrangement of
magnetic moments on the Fe1 and Fe2 sites, is centrosymmetric, and
thus the system is non-toroidic in the localized moment limit (see
also Ref.~\onlinecite{Arima_et_al:2004}). This is consistent with
Eq.~(\ref{GFO:t2}), which for $y_1 - y_2 = \frac{1}{4}l$ results only
in a trivial non-toroidal component of $\vec{T}$ along the $a$
direction. We point out that the non-toroidicity for
$y_1-y_2=\frac{1}{4}l$ holds true only for the case of localized
magnetic moments, where the presence of all nonmagnetic ions is
neglected. The full crystallographic symmetry of this system (given by
both magnetic and nonmagnetic ions) is non-centrosymmetric even for
$y_1 - y_2 = \frac{1}{4}l$. In fact, GaFeO$_3$ is an example of a
pyroelectric crystal that is polar but not ferroelectric, i.e. the
polarization cannot be switched, since it does not result from a small
distortion of a centrosymmetric reference structure. However, in the
localized moment picture the system is non-toroidal for $y_1 - y_2 =
\frac{1}{4}l$ and we can evaluate the spontaneous toroidization with
respect to this reference configuration. The relevant structural
parameters determined experimentally in
Ref.~\onlinecite{Arima_et_al:2004} at 4\,K are: $a=8.719$\,\AA,
$b=9.368$\,\AA, $c=5.067$\,\AA, $y_1=0.5831$, and $y_2=0.7998$. This
gives a spontaneous toroidization of $T_s =
\frac{2Sb}{\Omega}(0.25-0.2167) = 7.5 \cdot 10^{-3}
\mu_\text{B}/\textrm{\AA}^2$, corresponding to a spontaneous toroidal
moment per unit cell of $3.1 \mu_\text{B} \textrm{\AA}$. Here, we used
the formal magnetic moment $S=5\mu_\text{B}$ of Fe$^{3+}$.

It can be seen from Eq.~(\ref{GFO:t2}) that the set of toroidization
values is also centrosymmetric around the origin if $y_1 - y_2$ is
equal to any integer multiple of $\frac{1}{8}$, even though only for
$y_1-y_2 = \frac{1}{4}l$ the corresponding magnetic moment
configuration is non-toroidal as discussed in the previous
paragraph. This shows that even though the set of toroidization values
of a non-toroidal structure is always centrosymmetric, the converse is
not necessarily true. A centrosymmetric set of toroidization values
does not necessarily correspond to a non-toroidal state. Thus, for
$y_1 - y_2 = (2k+1) \cdot \frac{1}{8}$ the higher symmetry of the
toroidization values is accidental and does not correspond to a
vanishing macroscopic toroidization.

The toroidal moment of a single unit cell of GaFeO$_3$ was also
evaluated in Ref.~\onlinecite{Popov_et_al:1998}, without taking into
account the multivaluedness of the toroidization due to the periodic
boundary conditions. A value of $t_0 = 24.155\, \mu_B$\AA\ along the
$a$ direction was reported for the centrosymmetric reference structure
with $y_1 - y_2 = -0.25$, and the spontaneous toroidal moment was
specified as 0.03\,$t_0$. It is unclear from
Ref.~\onlinecite{Popov_et_al:1998} whether mixing of Fe ions onto the
Ga sites was included in the calculation and therefore a direct
comparison with our calculation is not possible.

\subsection{BiFeO$_3$}
\label{sec:BFO}

BiFeO$_3$ is a multiferroic material of high practical interest since
it combines both magnetic and ferroelectric order above room
temperature (see Ref.~\onlinecite{Wang_et_al:2003}). It exhibits a
rhombohedrally distorted perovskite structure (space group $R3c$)
involving both polar displacements of the ions along the [111]
direction and counter-rotations of oxygen octahedra around this
direction.\cite{Michel_et_al:1969,Kubel/Schmid:1990,Neaton_et_al:2005}
The spin structure of BiFeO$_3$ is a superposition of various
components. In a first approximation, the spins order in a G-type
antiferromagnetic structure, where all neighboring magnetic moments
are oriented antiparallel to each other.\cite{Fischer_et_al:1980} In
addition, in bulk BiFeO$_3$ the axis along which the spins are aligned
rotates throughout the crystal, leading to an additional spiral spin
structure with a large period of
$\sim$620\,\AA.\cite{Sosnowska/Peterlin-Neumaier/Streichele:1982}
However, this spiral component is absent in thin film
samples,\cite{Bea_et_al:2007} where instead a weak ferromagnetic
moment, resulting from a small canting of the magnetic moments, has
been reported.\cite{Wang_et_al:2003,Ederer/Spaldin:2006} Here, we
exclude the bulk spiral component as well as the small weakly
ferromagnetic component from the discussion. Depending on the
direction of the antiparallel Fe moments, the magnetic point group is
either $3m$ (if the moments are aligned along the polar $z$ axis), $m$
(if the moments are oriented perpendicular to the polar axis and
parallel to the glide plane), or $m'$ (with the moments perpendicular
to the glide plane); all of these symmetries allow for a macroscopic
toroidization. First-principles calculations showed that $3m$
symmetry, which would not allow for a macroscopic magnetization, is
energetically unfavorable.\cite{Ederer/Spaldin:2005}

BiFeO$_3$ provides an instructive example to illustrate the
limitations of the localized moment approach to toroidal moments based
on Eq.~(\ref{t_spin}). Both space and time-inversion symmetries are
broken in BiFeO$_3$, so that in principle a toroidal moment is
symmetry-allowed for this system. Indeed, an antisymmetric component
of the magneto-electric tensor, indicating a non-vanishing toroidal
moment, has been measured at high magnetic fields where the bulk
spiral spin structure is destroyed.\cite{Popov_et_al:2001} However, if
we consider only the localized spins on the Fe sites, the magnetic
lattice is centrosymmetric, i.e. it has a higher symmetry than the
full magnetization density $\vec{\mu}(\vec{r})$ (see the discussion
towards the end of Sec.~\ref{definitions}), and thus the corresponding
toroidal moment vanishes. In the $R3c$ crystal structure of BiFeO$_3$
the inversion symmetry is lifted by displacements of the different
ionic sub-lattices relative to each other. Since the local moment
picture neglects the presence of all nonmagnetic ions, this inversion
symmetry-breaking is not present in the purely magnetic lattice.

If we take the rhombohedral axis of BiFeO$_3$ as the $z$-direction
(hexagonal setup of the rhombohedral unit cell) and consider a perfect
G-type ordering with the magnetic moments of the Fe cations along $\pm
\hat{\vec{x}}$, then the calculated toroidization is:
\begin{equation}
\label{t_bfo}
\vec{T}_n = \frac{S}{2\Omega}\left( \begin{array}{c} 0 \\ \frac{c}{6}
   + n \frac{c}{3} \\ 0 \end{array} \right) \quad .
\end{equation}
Here, $c$ is the lattice constant of BiFeO$_3$ along $z$ within the
hexagonal setup. The first term along the $y$ direction in
Eq.~(\ref{t_bfo}) is the trivial half-toroidization increment, and the
whole set of toroidization values is centrosymmetric,
i.e. non-toroidal. This shows that it is not always sufficient to
consider magnetic dipole moments localized at the sites of the
magnetic cations, but that the spatial moment distribution can be very
important. The exact magnetic moment density $\vec{\mu}(\vec{r})$
always reflects the full magnetic space group symmetry of the system,
whereas the reduction to localized magnetic moments can result in a
higher symmetry than that of the full system.

\section{Summary, conclusions, and outlook}
\label{conclusions}

In summary we have presented a detailed study of magnetic toroidal
moments in bulk periodic solids in the limit where the toroidal moment
is caused by a time and space reversal symmetry breaking arrangement
of localized magnetic moments. We have reviewed the basic microscopic
definitions and showed that the periodic boundary conditions lead to a
multivaluedness of the toroidization, which suggests that only
differences in toroidization are well-defined observable quantities.
We suggest that the origin dependence of the toroidal moment should be
treated by decomposing the magnetic moment arrangement into a fully
compensated antiferromagnetic and an uncompensated ferromagnetic
component, so that only the ferromagnetic component depends on the
origin. Differences in toroidization resulting from the compensated
part of the moment configuration can be evaluated rather
straightforwardly, if one keeps in mind the macroscopic symmetry
properties of the system. We have illustrated the main concepts and
difficulties in evaluating magnetic toroidization in periodic systems
by first discussing the simple example of a one-dimensional
antiferromagnetic chain, and we have then analyzed the toroidization
for four example materials.

In addition to illustrating the general consequences of the origin
dependence and the multivaluedness of the toroidal moment in periodic
systems in the localized moment limit, our main conclusion is that it
is important to be aware of the macroscopic symmetry properties when
evaluating toroidization changes. This is particularly striking in the
example of the distorted one-dimensional antiferromagnetic chain
discussed in Sec.\ref{chain}, where the change in polarization due to
a structural distortion can be calculated straightforwardly, whereas
in the case of a magnetic moment reversal one has to subtract the
improper toroidization change that is caused by the corresponding
change in the toroidization increment. We have also shown the
limitations of the local moment picture in evaluating toroidal moments
in cases where the reduction to localized moments changes the symmetry
of the system.

An open question, which we have not addressed in our theoretical
analysis, is how the spontaneous toroidization can be measured
experimentally. According to the fundamental definition of the
toroidal moment, this is in principle possible by measuring the torque
on a sample that is placed in an inhomogeneous magnetic
field. However, either a field with a constant curl over the whole
dimension of the sample has to be generated, or the effects of other
multipole moments that couple to other field components have to be
subtracted appropriately. To our knowledge such a measurement has not
been attempted as of yet. So far, experimental evidence for toroidal
moments has been based mostly on the detection of an asymmetric
component of the magneto-electric tensor, but since the corresponding
prefactors are not known, an absolute quantitative determination of
$\vec{T}$ is not possible. Similarly, the nonlinear optical techniques
used in Ref.~\onlinecite{VanAken_et_al:2007} are mostly sensitive to
symmetry breaking, but are at best semi-quantitative. Overall it
appears that quantitative measurements of toroidal moments are a
challenging task, but in principle possible.

Finally, we comment on some aspects of toroidal moments that have been
discussed in a sometimes confusing way in the literature. First, the
relation between the toroidal moment and the asymmetric component of
the linear magneto-electric effect, and second the outer product of
the polarization with the magnetization, which has sometimes been
interpreted as a toroidal moment.

From a macroscopic symmetry point of view, the symmetries which allow
for a macroscopic toroidal moment are identical with that allowing for
an antisymmetric component of the linear magneto-electric effect
tensor. The relation between these two quantities can be seen by
analyzing the following free energy expression (see also
Ref.~\onlinecite{Sannikov:1997}):
\begin{multline}
\label{free-energy}
U = \frac{1}{2} \kappa P^2 - \vec{P}\cdot\vec{E} + \frac{1}{2} B M^2 -
\vec{M}\cdot\vec{H} \\ + \frac{1}{2} A T^2 + \frac{1}{4} C T^4 + a
\vec{T} \cdot \left( \vec{P} \times \vec{M} \right) \quad .
\end{multline}
This is the simplest possible free energy expression that can
simultaneously describe (i) a phase transition from a para-toroidic
($T=|\vec{T}|=0$) into a ferrotoroidic phase ($T \neq 0$), (ii) the
coupling of the electric polarization $\vec{P}$ and the magnetization
$\vec{M}$ to the electric field $\vec{E}$ and the magnetic field
$\vec{H}$, respectively, and (iii) a coupling between the electric
polarization, the magnetization, and the toroidization. In
Eq.~(\ref{free-energy}) $\kappa$ and $B$ are the inverse electric and
magnetic susceptibilities, $A$ and $C$ are temperature dependent
coefficients, and $a$ determines the strength of the magneto-electric
coupling. The trilinear form of the coupling term in
Eq.~(\ref{free-energy}) is the lowest possible order that is
compatible with the overall space and time reversal symmetries. The
equilibrium values for $\vec{P}$ and $\vec{M}$ can be obtained by
minimizing Eq.~(\ref{free-energy}). This leads to:
\begin{equation}
\label{pol}
\vec{P} = \frac{1}{\kappa} \left\{ \vec{E} - a ( \vec{M} \times
\vec{T} ) \right\}
\end{equation}
and
\begin{equation}
\label{mag}
\vec{M} = \frac{1}{B} \left\{ \vec{H} - a ( \vec{T} \times
\vec{P} ) \right\} \quad .
\end{equation}
If one inserts Eq.~(\ref{mag}) into Eq.~(\ref{pol}) one obtains (to
leading order in $\vec{T}$):
\begin{equation}
\label{PME}
\vec{P} = \frac{1}{\kappa} \vec{E}  - \frac{a}{\kappa B} ( \vec{H}
\times \vec{T} ) \quad .
\end{equation}
The last term in Eq.~(\ref{PME}) represents an antisymmetric linear
magneto-electric effect proportional to the toroidization. Thus, the
presence of the trilinear coupling term between toroidization,
magnetization, and polarization in Eq.~(\ref{free-energy}) gives rise
to an antisymmetric magneto-electric effect $\vec{P} = \alpha \vec{H}$
in the ferrotoroidic phase, with $\alpha_{ij} = -\frac{a}{\kappa B}
\sum_k \epsilon_{ijk} T_k$. Note that in general other terms in the
free-energy expansion can give rise to additional symmetric
contributions to the magneto-electric tensor, which are not
proportional to the toroidal moment.

In Eq.~(\ref{free-energy}) only the magnetization and the polarization
couple to $\vec{H}$ and $\vec{E}$, the toroidization in general does
not couple to any homogeneous external fields, in agreement with the
fundamental definitions discussed in Sec.~\ref{definitions} (in
particular Eq.~(\ref{Eoft})). The \emph{effective} coupling
represented by the invariant $E_\text{ME} \sim \vec{T} \cdot ( \vec{E}
\times \vec{H} )$ (discussed in
Refs.~\onlinecite{Dubovik/Tugushev:1990,Gorbatsevich/Kopaev:1994,Schmid:2003})
arises from the trilinear coupling term in Eq.~(\ref{free-energy}) if
$\vec{P}$ and $\vec{M}$ are substituted by their corresponding fields,
by inserting Eqs.~(\ref{pol}) and (\ref{mag}) into
Eq.~(\ref{free-energy}). Of course it depends on the problem at hand
whether it is more convenient to use the fields $\vec{E}$ and
$\vec{H}$ or the magnetization $\vec{M}$ and polarization $\vec{P}$ as
free variables. It is worth pointing out, though, that the two cases
should be carefully distinguished. One can either use a description
where $\vec{P}$, $\vec{T}$, and $\vec{M}$ are the free variables,
$\vec{P}$ and $\vec{M}$ couple linearly to their corresponding
fields, and there is a trilinear coupling term of the form $\vec{T}
\cdot (\vec{P} \times \vec{M})$, or one can alternatively use a
picture where the variables $\vec{P}$ and $\vec{M}$ are eliminated
altogether and are replaced by the field variables $\vec{E}$ and
$\vec{H}$. In this case the trilinear coupling term leads to an
effective coupling of $\vec{T}$ to $\vec{E} \times \vec{H}$.

Another source of confusion is the relation of the toroidization to
the cross product $\vec{P} \times \vec{M}$, which has the same time
and space reversal symmetry as $\vec{T}$. This has led to a number of
instances in the literature in which $\vec{P} \times \vec{M}$ itself
has been described as a ``toroidal
moment''.\cite{Arima_et_al:2005,Sawada/Nagaosa:2005,Yamasaki_et_al:2006}
We point out that this is generally not correct. From our discussion
in Sec.~\ref{origin} it becomes clear that $\vec{P} \times \vec{M}$
can never describe a toroidal moment resulting from the compensated
part of a magnetic moment configuration (for which $\vec{M} = 0$). We
have also shown in Sec.~\ref{origin} that the change in toroidal
moment resulting from the uncompensated part of the magnetic moment
configuration can be expressed as $\Delta \tilde{\vec{t}} = 1/2 \Delta
\bar{\vec{R}} \times \vec{m}$, where $\Delta \bar{\vec{R}}$ is the
average (non-centrosymmetric) displacement of the magnetic moments. We
point out that, at least in the localized moment picture, $\Delta
\bar{\vec{R}}$ is in general not proportional to $\vec{P}$, and
therefore the toroidal moment is in general not proportional to
$\vec{P} \times \vec{M}$. In most magnetic ferroelectrics, the
polarization is related to a rigid shift of the magnetic cations
relative to the non-magnetic ions, which does not affect the average
position $\bar{\vec{R}}$ of the magnetic cations. It is therefore not
clear whether the conditions for $\vec{T} \sim \vec{P} \times \vec{M}$
are fulfilled in any currently known material.

It has been argued that ferrotoroidicity is a key concept for fitting
all forms of ferroic order in a simple fundamental scheme based on the
different transformation properties of the corresponding order
parameters with respect to time and space inversion (see
Refs.~\onlinecite{Schmid:2003,Schmid:2004,VanAken_et_al:2007}, in
particular Fig.~2 in Ref.~\onlinecite{VanAken_et_al:2007}). The four
fundamental forms of ferroic order listed in these references are
ferroelasticity, ferroelectricity, ferromagnetism, and
ferrotoroidicity, with order parameters transforming according to the
four different representations of the ``parity group'', which is
generated by the two operations of time and space
reversal.\cite{Ascher:1974} A similar scheme has also been proposed in
Ref.~\onlinecite{Dubovik/Tugushev:1990}, but with the electric
toroidal moment $\vec{g}$ (see Sec.~\ref{intro} and also
Refs.~\onlinecite{Naumov/Bellaiche/Fu:2004,Prosandeev_et_al:2006}) as
the time and space symmetric order parameter instead of the
ferroelastic strain tensor. The latter classification scheme seems
more natural to us, since in this case all ferroic order parameters
are vector quantities. It is thus very important to clearly
distinguish between \emph{magnetic} ferrotoroidicity and
\emph{electric} ferrotoroidicity, which correspond to different
representations of the parity group. On the other hand, the existence
of ferroelasticity, with the second-rank strain tensor as time and
space invariant order parameter, raises the question, whether other
ferroic second-rank tensor order parameters, corresponding to
different representations of the parity group, can be identified in
the future.

\begin{acknowledgments}
This work was supported by the Division of Materials Research of the
National Science Foundation (NSF), under grant number DMR-0605852 and
by the MRSEC Program of the NSF under award number DMR-0213574. The
authors thank Manfred Fiebig and Hans Schmid for useful
discussions. N.S. thanks the Miller Institute at UC Berkeley for their
support through a Miller Research Professorship.
\end{acknowledgments}

\bibliography{references}

\begin{thebibliography}{54}
\bibitem[{foo({\natexlab{c}})}]{footnote:compensated}
\bibinfo{note}{We point out that in the present paper ``compensated'' and
  ``uncompensated'' always refers to the decomposition defined above
  Eq.~(\ref{t_comp}), i.e. for the purpose of our discussion ``uncompensated''
  always means that all magnetic moments are perfectly collinear and have
  exactly the same magnitude.}
\bibitem[{foo({\natexlab{d}})}]{footnote:rbar}
\bibinfo{note}{Note that $\bar{\vec{R}}$ is defined only in terms of the
  positions of the magnetic cations, not from the full crystallographic
  structure}.
\end{thebibliography}

\end{document}